\documentclass[fleqn,usenatbib]{mnras}

\usepackage{newtxtext,newtxmath}

\usepackage[T1]{fontenc}

\DeclareRobustCommand{\VAN}[3]{#2}
\let\VANthebibliography\thebibliography
\def\thebibliography{\DeclareRobustCommand{\VAN}[3]{##3}\VANthebibliography}

\usepackage{graphicx}	
\usepackage{amsmath}	

\usepackage[normalem]{ulem}

\newcommand{\msun}{\mbox{M$_{\odot}$}}
\newcommand{\rsun}{R$_{\odot}$}

\newcommand{\fss}{\hbox{$.\!\!^\mathrm{s}$}}        
\newcommand{\h}{$^\mathrm{h}$}
\newcommand{\m}{$^\mathrm{m}$}

\newcommand{\gr}{$\gamma$-ray}
\newcommand{\swift}{\textit{Swift}}
\newcommand{\eros}{\textit{eROSITA}}
\newcommand{\src}{J2054}
\newcommand{\sxps}{2SXPS~J205359.0+690518}
\newcommand{\fgl}{4FGL~J2054.2+6904}
\newcommand{\pso}{PSO~J313.4958+69.0888}
\newcommand{\ergs}{erg~s$^{-1}$}
\newcommand{\flux}{erg~s$^{-1}$~cm$^{-2}$}

\newcommand{\degs}{\ifmmode ^{\circ}\else$^{\circ}$\fi}
\newcommand{\amin}{\ifmmode ^{\prime}\else$^{\prime}$\fi}
\def\asec{\ifmmode ^{\prime\prime}\else$^{\prime\prime}$\fi}

\newcommand{\gaia}{\textit{Gaia}}
\newcommand{\ps}{Pan-STARRS}

\newcommand{\fermi}{\textit{Fermi}}

\usepackage{threeparttable}

\title[A new redback pulsar candidate \fgl]{A new redback pulsar candidate  \fgl}

\author[A. V. Karpova et al.]{
A. V. Karpova,$^{1}$\thanks{E-mail: annakarpova1989@gmail.com} 
D. A. Zyuzin,$^{1}$
Yu. A. Shibanov,$^{1}$
M. R. Gilfanov$^{2,3}$
\\
$^{1}$Ioffe Institute, Politekhnicheskaya 26, St. Petersburg 194021, Russia\\
$^{2}$Space Research Institute, Russian Academy of Sciences, Profsoyuznaya 84/32, Moscow 117997 Russia \\
$^{3}$Max-Planck-Institut f\"ur Astrophysik, Karl-Schwarzschild-Str. 1, D-85741 Garching, Germany 
}

\date{Accepted XXX. Received YYY; in original form ZZZ}

\pubyear{2023}

\begin{document}
\label{firstpage}
\pagerange{\pageref{firstpage}--\pageref{lastpage}}
\maketitle

\begin{abstract}
The \fermi\ catalogue contains about 2000 unassociated $\gamma$-ray sources. 
Some of them were recently identified as pulsars,  
including so called redbacks and black widows, which are  millisecond 
pulsars in tight  binary systems with    non- and partially-degenerate  low-mass  stellar  
companions irradiated by the pulsar wind.
We study a 
likely optical and X-ray 
counterpart of the \fermi\ source 4FGL J2054.2$+$6904  proposed earlier as a pulsar candidate. 
We use archival optical data as well as \swift/XRT\ 
and SRG/\eros\ X-ray  data to clarify  
its nature. Using Zwicky Transient Facility  data  in  $g$ and $r$ bands spanning over 4.7 years,  
 we find a  period  of $\approx$7.5 h. The folded light curve has a smooth sinusoidal shape with the peak-to-peak amplitude of $\approx$0.4 mag. 
The spectral fit to the optical spectral energy distribution of the counterpart candidate 
gives the star radius 
of 0.5$\pm$0.1~\rsun\ and
temperature of 5500$\pm$300 K implying a  G2--G9-type star. 
Its X-ray spectrum is well fitted by
an absorbed power law with the photon index of 1.0$\pm$0.3
and unabsorbed flux of $\approx 2\times10^{-13}$ \flux.  
All the properties of 4FGL J2054.2$+$6904 and its presumed counterpart 
suggest that it 
is a member of the redback family.  
\end{abstract}




\section{Introduction}

Spider millisecond pulsars (MSPs) represent a class of pulsar binaries 
with low-mass companions and tight orbital periods of $\lesssim 1$ d 
\citep{roberts2013}. The side of the companion star facing the pulsar 
is illuminated, heated and evaporated by the pulsar wind. The ablated material often 
lead to extensive eclipses of the pulsar emission. Spider pulsars are 
divided into redbacks (RBs) and black widows (BWs). The former has non-degenerate
companions with masses $M_{\rm c}=$~0.1--1~\msun\ while companions of the latter
are very low-mass, $M_{\rm c} \lesssim 0.05$~\msun, partially degenerate stars
\citep{roberts2013}. 
Studying spider systems can provide
information about the mass of the hosted neutron star (NS), which in such systems can exceed
2~\msun\ \citep[e.g.][]{romani2022}. 
This is important for constraining the equation of 
state of the superdense matter inside NSs.

To date, about 40 
RBs and 70 BWs has been discovered. A half of them 
are located in the Galactic disc \citep{swihart2022,strader2019} and a half -- 
in globular clusters\footnote{See \url{http://www.naic.edu/~pfreire/GCpsr.html}}.
Among them, there are three RBs, PSRs J1023+0038, J1824$-$2452I and J1227$-$4853, that show 
transitions between accretion-powered and rotation-powered pulsar states
\citep{archibald2009,papitto2013,bassa2014,roy2015}.
This provides a link between MSPs and low-mass X-ray binaries (LMXBs) 
supporting the recycling scenario according to which MSPs are spun-up 
due to accretion of matter from the secondary star \citep{Bisnovatyi-Kogan1974,alpar1982}. 
Nevertheless, there are still many unanswered questions about details of formation 
and evolution of spider systems 
\citep[e.g.][]{chen2013,benvenuto2014,ablimit2019,ginzburg&quataert2021,guo2022}. 

A great success in uncovering spider pulsars has been achieved thanks 
to the \fermi\ Gamma-ray Space Telescope. About two tens of promising
RB and BW candidates are proposed through identifications of possible 
optical and/or X-ray counterparts to $\gamma$-ray sources
\citep{strader2019,strader2021,miller2020,swihart2020,swihart2021,swihart2022,au2023}. 
However, the Data Release 3 of the fourth \fermi\ Large Area Telescope (LAT) 
catalog (4FGL-DR3) contains more than 2000 unassociated $\gamma$-ray sources
\citep{4fgl-dr3}. Some of them were classified as binary  pulsar candidates
\citep{Kerby2021}. Among them is \fgl\ (hereafter \src). Its flux
in the 0.1--100 GeV band is $(4.2 \pm 0.5)\times 10^{-12}$ \flux\ and
its \gr\ spectrum can be fitted with a logParabola model \citep{4fgl-dr3}.
Its possible X-ray  and  optical counterpart was found with \swift\ \citep{Kerby2021}. 
In the second \textit{Swift} X-ray Point Source (2SXPS) catalog \citep{2swift} it is designated
as 2SXPS~J205359.0+690518 with coordinates 
R.A. = 20\h53\m59\fss10 and Dec. = +69\degs05\amin18\farcs5.

To better understand the nature of \src\ and its putative optical and X-ray  counterparts,     
we used various archival multi-epoch optical observations
and X-ray data from \swift/XRT, as well as from \eros\
\citep{erosita2021} aboard the Spectrum-RG 
(SRG) orbital observatory \citep{Sunyaev2021}. 
Here we present analysis of these data.
Optical data are described in Sec.~\ref{sec:oi}, their  timing analysis -- in Sec.~\ref{sec:lc} and optical spectrum is studied in Sec.~\ref{sec:opt-sed}. 
The description of the X-ray data is presented 
in Sec.~\ref{sec:x-ray}. We discuss and summarise the results in Sec.~\ref{sec:discussion}.

 
\section{X-ray and optical data}
\label{sec:oi}


\begin{figure}
\begin{minipage}[h]{1\linewidth}
\center{\includegraphics[width=1\linewidth, trim={0 0 0 0.4cm},clip]{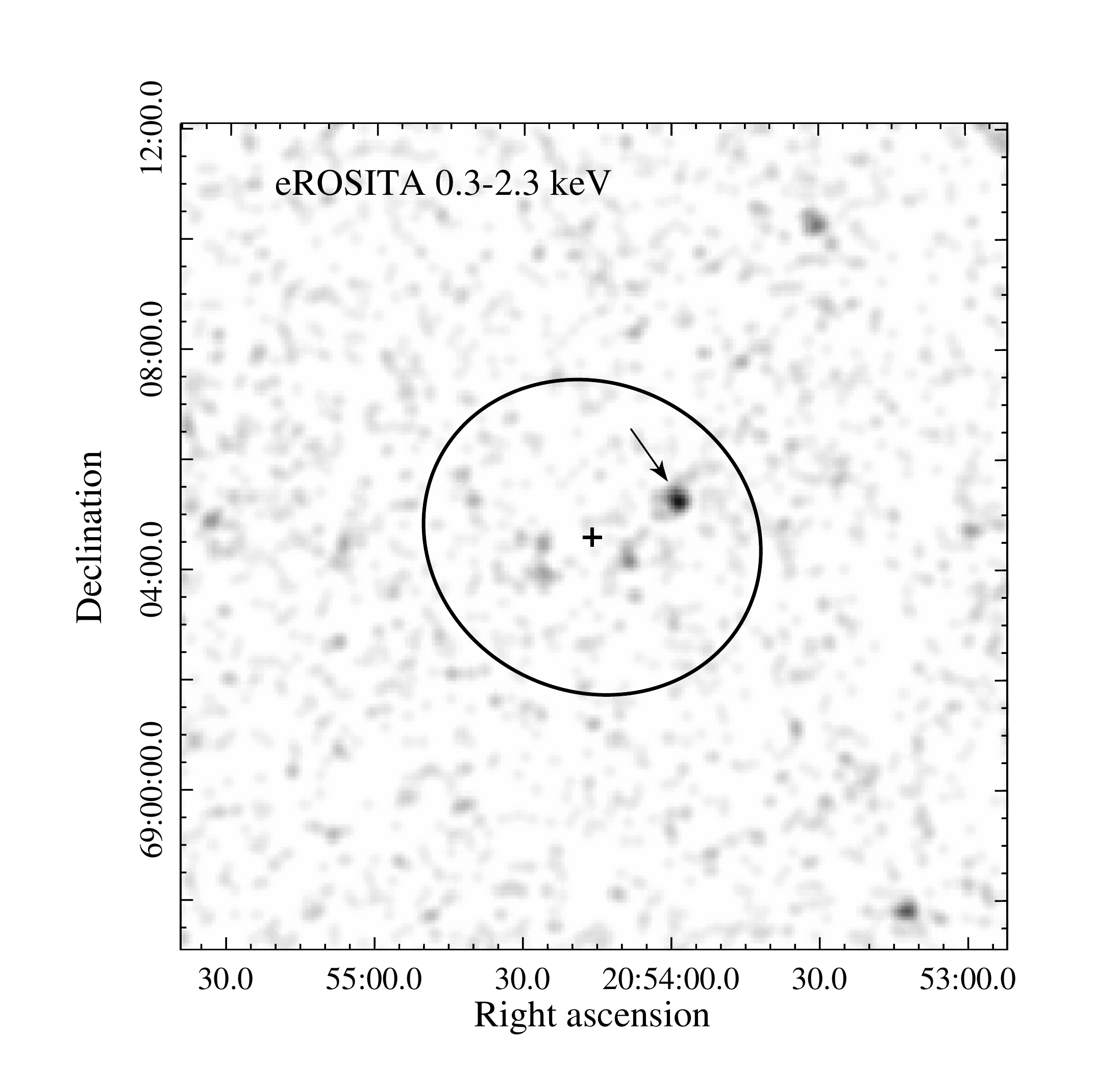}}
\end{minipage}
\begin{minipage}[h]{1.\linewidth}
\center{\includegraphics[width=1.0\linewidth,trim={0 0 0.7cm 0.3cm},clip]{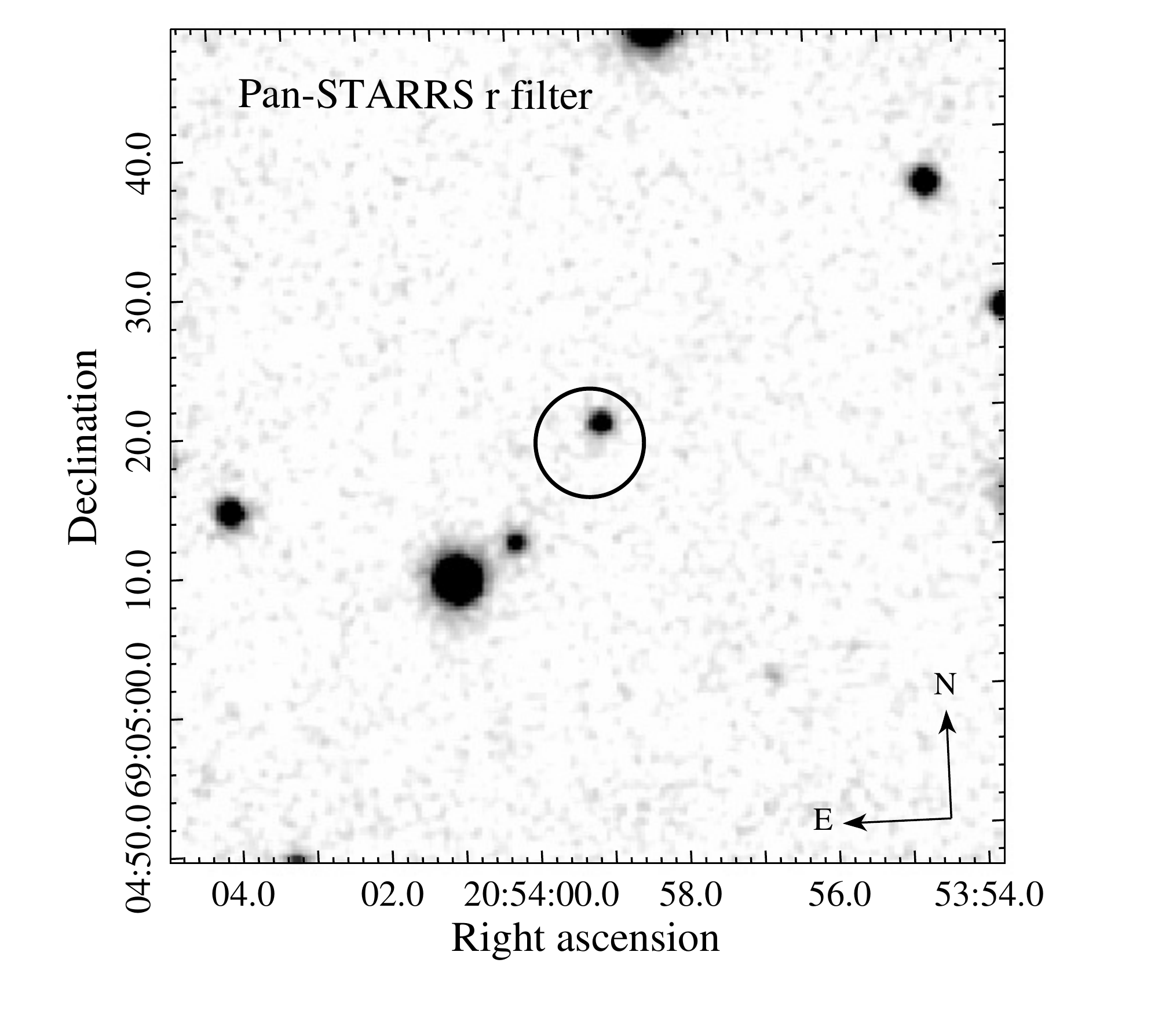}}
\end{minipage}
\caption{ \textit{Top}: SRG/\eros\ image of the \fgl\ field in the 0.3--2.3 keV band. The black 
ellipse shows 95 per cent uncertainty of the 
$\gamma$-ray source position marked by the black cross.
The proposed counterpart of \fgl, \sxps, is marked by the arrow. \textit{Bottom}: 1$\times$1 arcmin$^2$ \ps\ image of the \sxps\ field 
obtained in the $r$ band. The black circle shows the 90 per cent uncertainty of 
the X-ray position of the source taken from 2SXPS catalogue.
The compass corresponds to the equatorial system.}
\label{fig:xray+opt}
\end{figure}


\begin{table}
\renewcommand{\arraystretch}{1.2}
\caption{Parameters of the \src\ and its X-ray/optical counterpart candidate.}
\label{tab:pars}
\begin{center}
\begin{tabular}{lc}
\hline
\multicolumn{2}{c}{Parameters from the catalogues}                            \\  
R.A. (J2000)                                                    & 20\h53\m58\fs99298(7)          \\
Dec. (J2000)                                                    & +69\degs05\amin19\asec.7063(5) \\
Galactic longitude $l$, deg                                     & 104.365                           \\
Galactic latitude $b$, deg                                      & 15.312                            \\
P.m. in R.A. direction $\mu_{\alpha}$cos$\delta$, mas yr$^{-1}$ & $-$5.4(5)                    \\
P.m. in Dec. direction $\mu_\delta$, mas yr$^{-1}$              & $-$5.5(5)                     \\
Distance $D$, kpc                                               & 1.5--5.3                          \\
$\gamma$-ray flux $F_\gamma$, $10^{-12}$ \flux                  & $4.2(5)$                        \\
Effective temperature $T_{\rm eff}^*$, K                        & $5820\pm412$ \\
Reddening $E(B-V)^*$, mag                                       & $0.35\pm0.04$ \\
\hline
\multicolumn{2}{c}{Parameters derived in this paper}                            \\
Orbital period $P_b$, h                                         & 7.4666(9)                         \\
 Maximum reddening $E(B-V)^\dag$, mag                           & 0.37 \\
Effective temperature $T_{\rm eff}$, K                          & $5500\pm300$  \\
X-ray flux $F_X$, $10^{-13}$ \flux\                             & $1.7^{+0.5}_{-0.3}$       \\ 
X-ray Luminosity $L_X$, $10^{32}$ \ergs\                        & 0.4--7.4 \\
$\gamma$-ray luminosity $L_\gamma$, $10^{34}$ \ergs\            & 0.1--1.6 \\
\hline
\end{tabular}
\end{center}
\begin{tablenotes}
\item Hereafter numbers in parentheses denote 1$\sigma$ uncertainties relating to the last 
significant digit quoted.
\item The coordinates are obtained from the \gaia\ catalogue as well as the proper motion (p.m.). 
\item $^*$Parameters from the TIC catalogue.
\item $^\dag$Maximum reddening in the source direction according to the dust map of \citet{dustmap2019}.
\item $F_X$ is the unabsorbed flux in the 0.5--10 keV range and $F_\gamma$ is the flux in the 0.1--100 GeV range. 
\end{tablenotes}
\end{table}

\begin{table}
\caption{Magnitudes of the counterpart candidate obtained from catalogues.}
\label{tab:mags}
\begin{center}
\begin{tabular}{ccccc}
\hline
\multicolumn{5}{c}{\ps\ (AB system), ID 19090313495683735}                                \\
$g$          & $r$          & $i$          & $z$         & $y$      \\   
20.82(3)     & 20.12(3)     & 19.71(4)     & 19.56(3)    & 19.37(4) \\
\hline
\multicolumn{5}{c}{\gaia, ID 2271107409667918080}                                           \\
             & $G$          & $G_{BP}$     & $G_{RP}$    &          \\
             & 20.142(9)    & 20.96(11)    & 19.50(8)    &          \\
\hline
\multicolumn{5}{c}{TIC, ID 1981395694}                                             \\
             &              & $T$          &             &          \\
             &              & 19.55(3)     &             &          \\
\hline
\multicolumn{5}{c}{UVOTSSC (AB system), ID 5624088}                             \\
             & $U$          & $B$          & $V$         &         \\
             & 23.23(15)    & 21.77(16)    & 20.52(12)   &         \\
\hline
\end{tabular}
\begin{tablenotes} ID is an object identifier in the catalogue.  
\end{tablenotes}
\end{center}
    
\end{table}
 
The \src\  was observed by SRG/\eros\ in the course of 5 all-sky surveys in 2020--2022 with
the total exposure time of 5.6 ks (vignetting corrected exposure of $\approx 2.7$ ks). 
The \eros\ raw data were processed by the calibration pipeline at IKI based on the \eros\ Science Analysis Software System (eSASS) and using pre- and in-flight calibration data.
The  image of the \src\ field is presented in the top panel of Fig.~\ref{fig:xray+opt},
where the likely X-ray counterpart is marked. The \eros\ coordinates of the latter are in agreement with the \sxps\ ones, obtained with \swift.

\citet{Kerby2021} identified the possible optical
counterpart of \src\ coinciding by position with \sxps\ using only \swift/UVOT data.
Investigating deeper images from the Panoramic Survey Telescope
and Rapid Response System survey (\ps, \citealt{ps2020}), 
we confirmed that there is only one optical source, \pso, within
the position uncertainty circle of \sxps\ (see Fig.~\ref{fig:xray+opt}, bottom).
Its parameters are shown in Table~\ref{tab:pars} and magnitudes 
-- in Table~\ref{tab:mags}.
There are  indications  of the source variability of about 0.3 mag
but the small number of observations did not allow us to
perform any informative timing analysis using \ps\ data.  

We also identified \pso\ in the \gaia\ catalogue \citep{gaia2016,gaia2022_dr3}, 
Transiting Exoplanet Survey Satellite (TESS) Input Catalog (TIC v.8.2, \citealt{tess8.2})
(see Table~\ref{tab:mags}) and Zwicky Transient Facility (ZTF, \citealt{ztf2019}) catalogue. 
The source parallax of 0.27$\pm$0.44 mas  is still  poorly defined by \gaia.   
Nevertheless, 
\citet{bailer-johnes2021} based on the \gaia\ data provide two different estimates of the  distance: in their terminology, one is the geometric distance  $D_{\rm geom} = 2.4^{+1.2}_{-0.9}$ kpc, and the other  is the   photogeometric distance $D_{\rm pgeom} = 4.7^{+0.6}_{-0.6}$ kpc.
In the following, to
estimate the parameters of the source, we will use the  distance range of 1.5--5.3 kpc  covering both estimates with their uncertainties.
The proper motion of the source is rather small (see Table~\ref{tab:pars}),
$\mu=7.7\pm0.5$ mas~yr$^{-1}$,  which corresponds to a transverse velocity
of about 50--200 km~s$^{-1}$ for the accepted  distance range. 
The ZTF data confirms the \pso\ variability and contains
enough measurements to search for periodicity. 

  
\section{Optical light curves and the orbital period}
\label{sec:lc}

\begin{figure}
\includegraphics[width=\columnwidth]{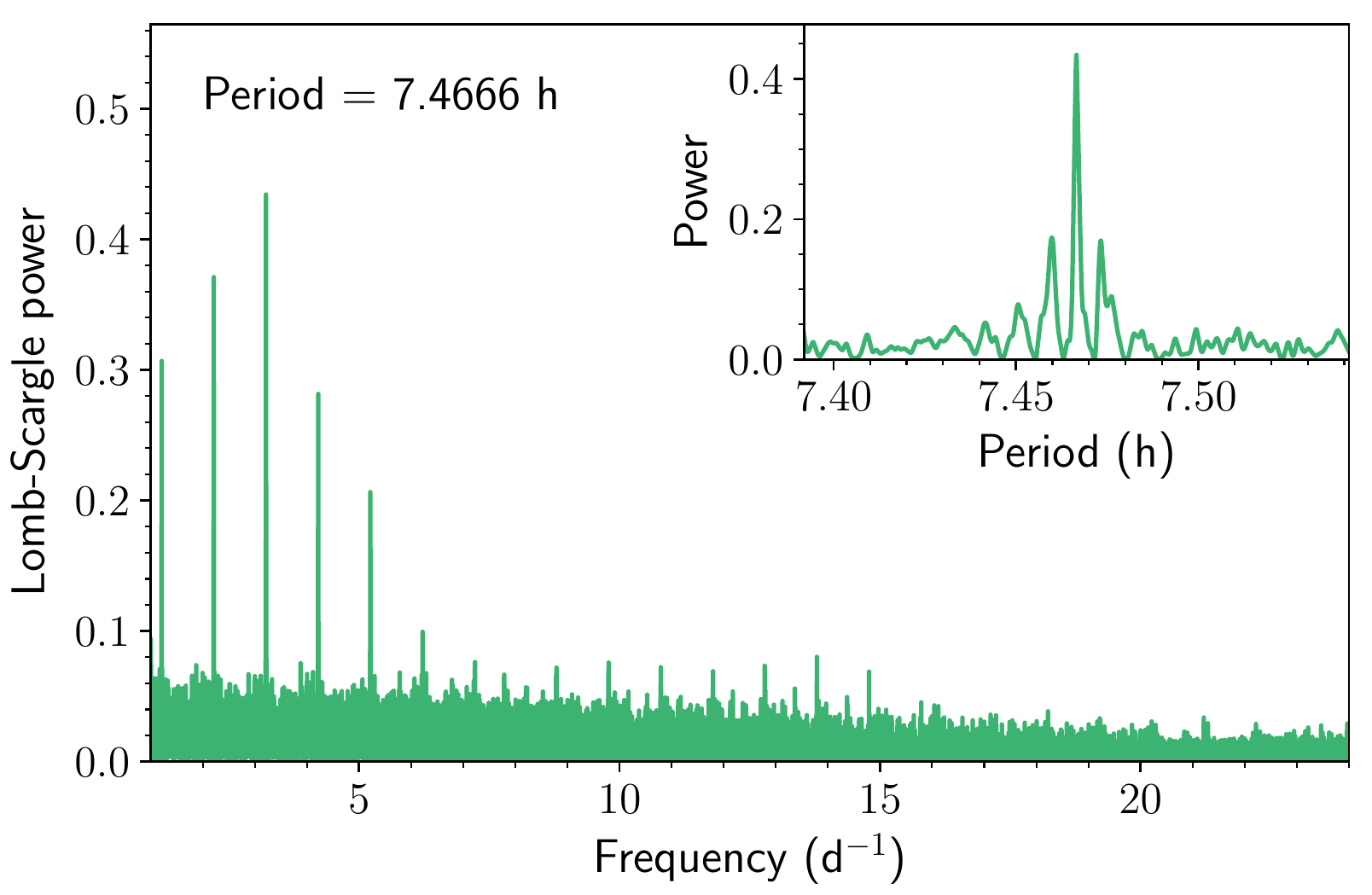} 
\caption{Lomb-Scargle periodogram  of the \src\ light curve obtained 
using the ZTF data in the $r$ band. The most likely period corresponding to the largest power peak is shown in the plot and the peak is enlarged in the inset. }
\label{fig:ztf-period}
\end{figure}

The ZTF DR16 archive  
contains  about six hundreds  brightness 
measurements  of the \pso\  in the $r$ band 
covering about 4.7 yr (MJD 58217--59937). We used 
the Lomb-Scargle method \citep{lomb1976,scargle1982}, to search for  periodic brightness 
variation in the range of 1--24 h. The resulting  
periodogram is plotted in Fig.~\ref{fig:ztf-period}. 
It shows a number of highly statistically significant peaks   caused by aliasing effects due to a 1 day windowing of ground-based ZTF observations \citep[see, e.g.,][]{VanderPlas}. 
The peaks are equally spaced with 
$\Delta f=1/24$ h$^{-1}$ around the main peak  
$f_{\rm ph}=1/P_{\rm ph}$: 
$f_n = f_{\rm ph} \pm n \Delta f$,  where $n$ is an  integer. The highest amplitude has the peak at the period  
$P_{\rm ph}=7.4666(9)$~h. Its 1$\sigma$  uncertainty quoted in the parentheses  is calculated as 
the half width at half maximum of the peak. The signal to noise ratio $SNR$ of the peak, 
computed using scripts developed by \citet{snad-ztf} for ZTF data\footnote{\url{https://ztf.snad.space/}}, is $\approx 40$.
The amplitude of any of the peaks of the lower amplitude is at least 
by $\ga 6\sigma$ lower. We therefore conclude that the $P_{\rm ph}=7.4666(9)$~h is the true period.

We also searched for periodicity in the ZTF data in the $g$ band. 
Though the number of measurements is about 2.5 times smaller than in the $r$ band
and the power spectrum is more noisy, we found a statisitcally significant peak  at the same period  $P_{\rm ph}$. 

\begin{figure}
\includegraphics[width=\columnwidth]{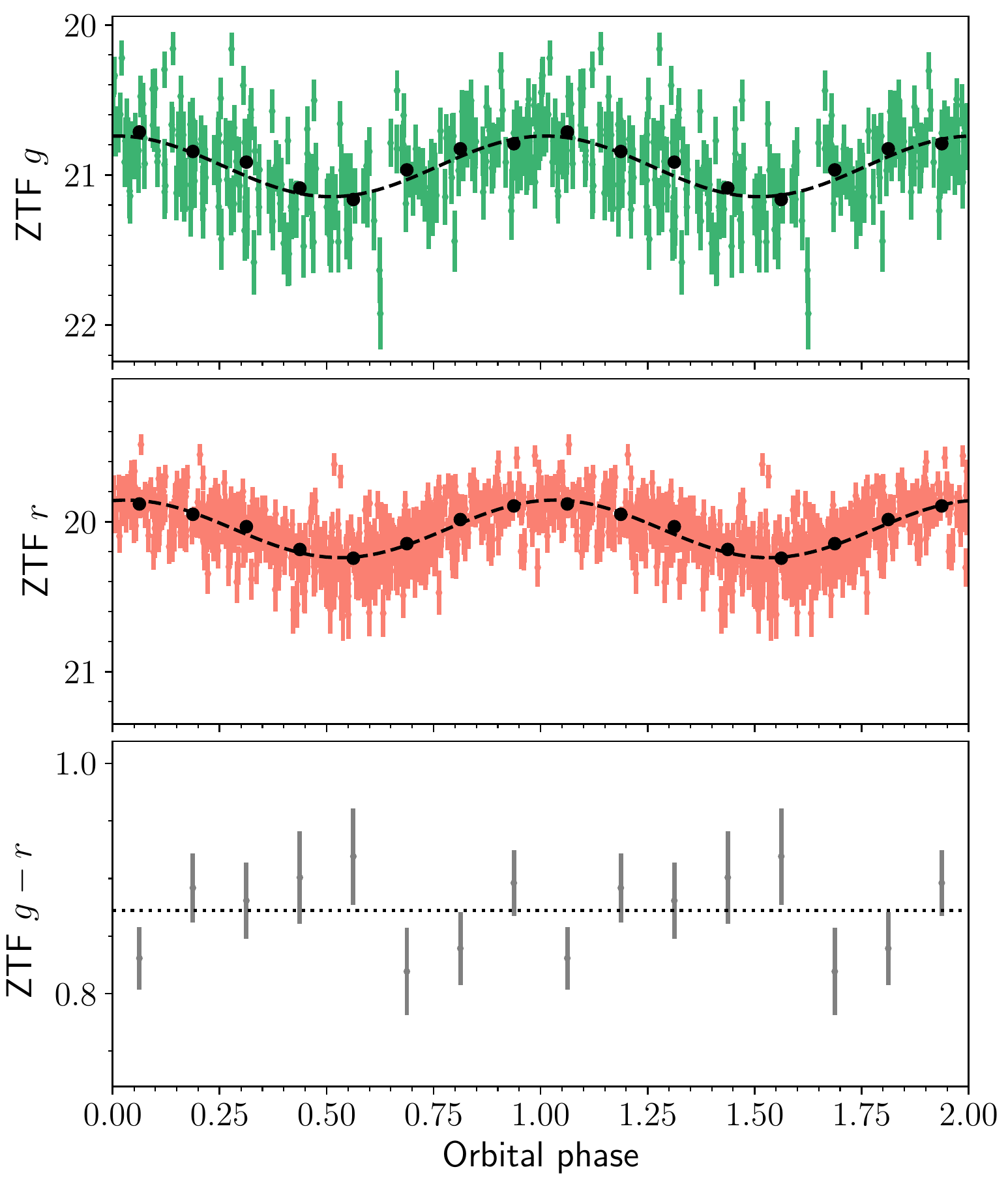}
\caption{Light curves of the \src\ optical counterpart candidate 
obtained using the ZTF data in the $g$ (\textit{top}) and $r$ bands (\textit{middle}) folded with the presumed  orbital period of 7.4666~h revealed in Fig.~\ref{fig:ztf-period}. The best-fitting sine curves are shown by the black dashed lines. The mean light curves averaged using eight phase bins are marked by the black circles. In the bottom panel, the $g-r$ colour curve calculated from the mean light curves is shown. The dotted line is the best-fitting constant function. }
    \label{fig:lc}
\end{figure}

The ZTF light curves in $g$ and $r$ bands folded 
with  
the period  $P_{\rm ph}$ 
are shown in the top and middle panels of Fig.~\ref{fig:lc}. They have a roughly sinusoidal form with 
a single broad peak per period. 
To estimate the amplitudes of brightness variations,
we fitted the folded light curves with functions of the form 
$A{\rm sin}(2\pi(\phi + \phi_0))+C$, where $\phi$ is the orbital phase and $C$ is a constant. As a result, 
the peak-to-peak amplitude (the difference between the maximum and minimum values, i.e. $2A$) is about  0.4 mag for both bands (see Fig.~\ref{fig:lc}).
We thus concluded that the \src\ is a binary system 
with the likely orbital period of $P_b = 7.4666(9)$~h.
We also binned the light curves in both bands and calculated the $g-r$ colour curve. The latter is presented in the bottom panel of Fig.~\ref{fig:lc}. No statistically significant variations of the colour
are seen. We fitted the colour with a constant function and obtained the mean colour of about 0.9 and the reduced $\chi^2_\nu=1.2$ for
7 degrees of freedom.

\section{Optical spectral energy distribution}
\label{sec:opt-sed}

We constructed  the spectral energy distribution (SED) of \pso\ using
magnitude measurements from the \ps, \gaia, TESS and UVOT data
(see Table~\ref{tab:mags} 
). 
The latter were obtained from the \swift/UVOT 
Serendipitous Source Catalog (UVOTSSC, 
\citealt{yershov2014,page2014}). 

\begin{figure}
\includegraphics[width=\columnwidth]{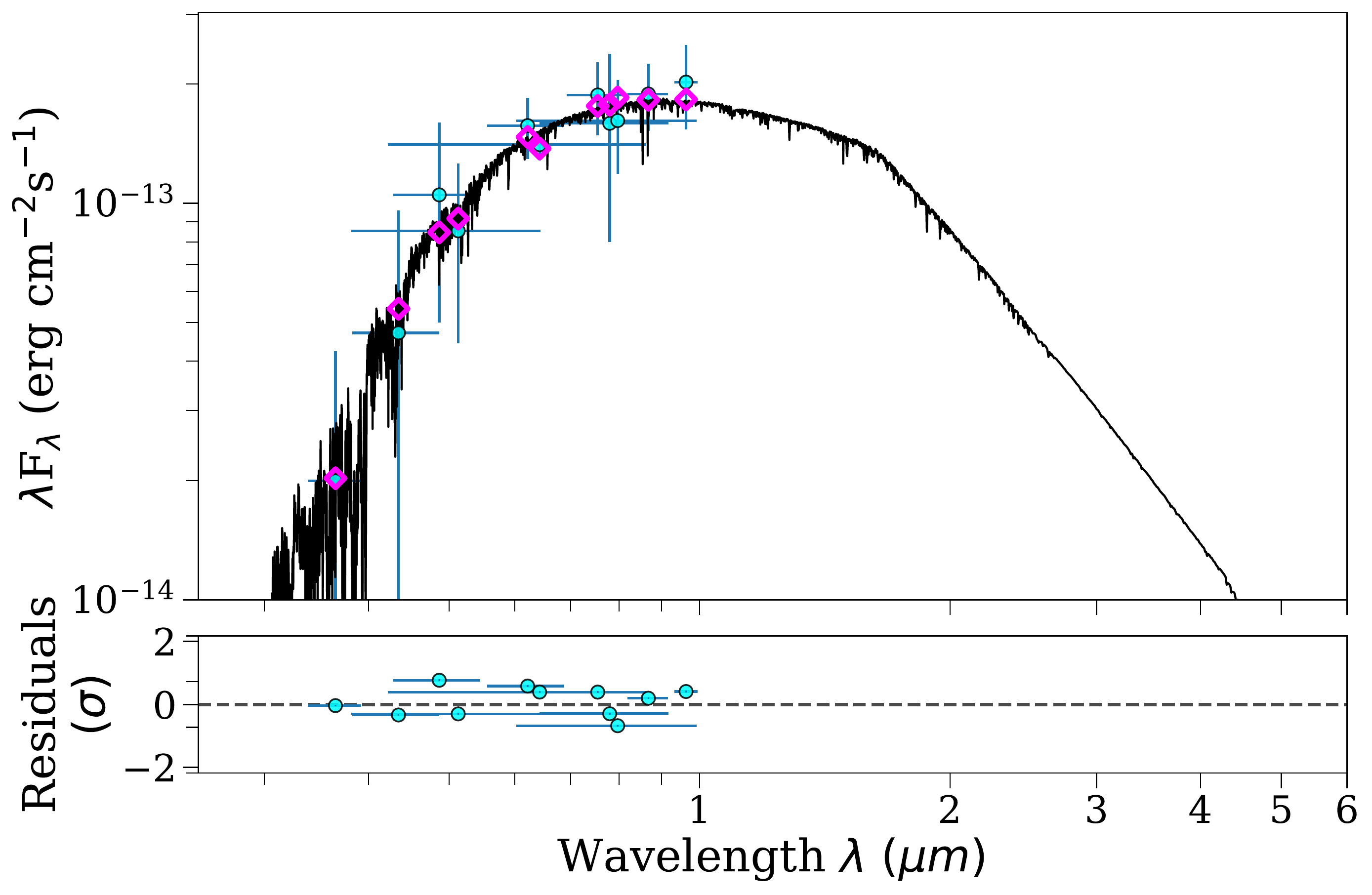}
\caption{ 
Spectral energy distribution  of the \src\ optical counterpart candidate (photometric data, circles with error bars),
the best-fitting model (black line) and fitting residuals normalized to the photometry errors (lower panel). 
Diamonds show the model broad-band fluxes.
}
\label{fig:sed}
\end{figure}

We fitted the SED with the spectral energy distribution 
Bayesian model averaging fitter ({\sc ariadne}) package \citep{ARIADNE}.
The model has the following parameters: reddening, distance,
surface gravity, radius, temperature and metallicity.
According to the 3D dust map of \citet{dustmap2019} 
the reddening $E(B-V)$ in the \pso\ direction
reaches its maximum value of 0.37\footnote{We note, that this value is in agreement with $E(B-V)$ provided in the TIC catalogue (see Table~\ref{tab:pars}).} at about 1.2~kpc
while the source is located at the larger distance, in the  range of 1.5--5.3~kpc. 
Thus, we fixed $E(B-V)$ at this value.  
Applying  a uniform prior for the distance, 
we obtained the best-fitting effective stellar 
temperature $T_{\rm eff}$~=~5500$\pm$300~K and  
radius $R$~=~0.5$\pm$0.1~\rsun. 
The best-fitting model is plotted in the top panel 
of Fig.~\ref{fig:sed}. The temperature 
is compatible with 5820$\pm$412~K obtained
by \citet{tess8.2}. In the absence of spectral data, it is impossible to   derive the metallicity and the gravitational mass of the star. 
We also obtained the distance $D$ of 3.7$\pm$0.7 kpc 
which is in agreement with the distance range provided by \gaia.
Nevertheless, we used the whole range mentioned above to calculate \src\ luminosities
to obtain most conservative estimates.

\section{\src\ properties in X-rays}
\label{sec:x-ray}

The X-ray spectrum of the \src\ counterpart candidate was studies
using archival \swift/XRT data and SRG/\eros\  all-sky survey data. 

The \swift\ spectrum 
was extracted with the \swift-XRT data products 
generator\footnote{\url{https://www.swift.ac.uk/user\_objects/}} \citep{evans2009}. 
We used observations carried out in 2009 and 2015. This resulted in 
the total effective exposure time of about 21 ks with 32 counts from the source
in the 0.3--10 keV band. The spectrum was grouped to ensure at least 1 count 
per energy bin.

The \eros\ source spectra were extracted in a circular region with a radius of 60 arcsec centered at the source position. An annulus region with the inner and outer radii of 150 and 300 arcsec around the source was used for the background extraction. As the source is fairly faint, we used the data of all 7 telescope modules.  No statistically significant  variability was detected between five individual sky surveys, although the statistical accuracy achieved in an individual survey is rather low, with typically $\sim$10--15 source counts registered in each  surveys. About 64 source counts were obtained in total.
The spectrum was grouped to ensure at least 3 counts per energy bin. 

\begin{figure}
\includegraphics[width=\columnwidth]{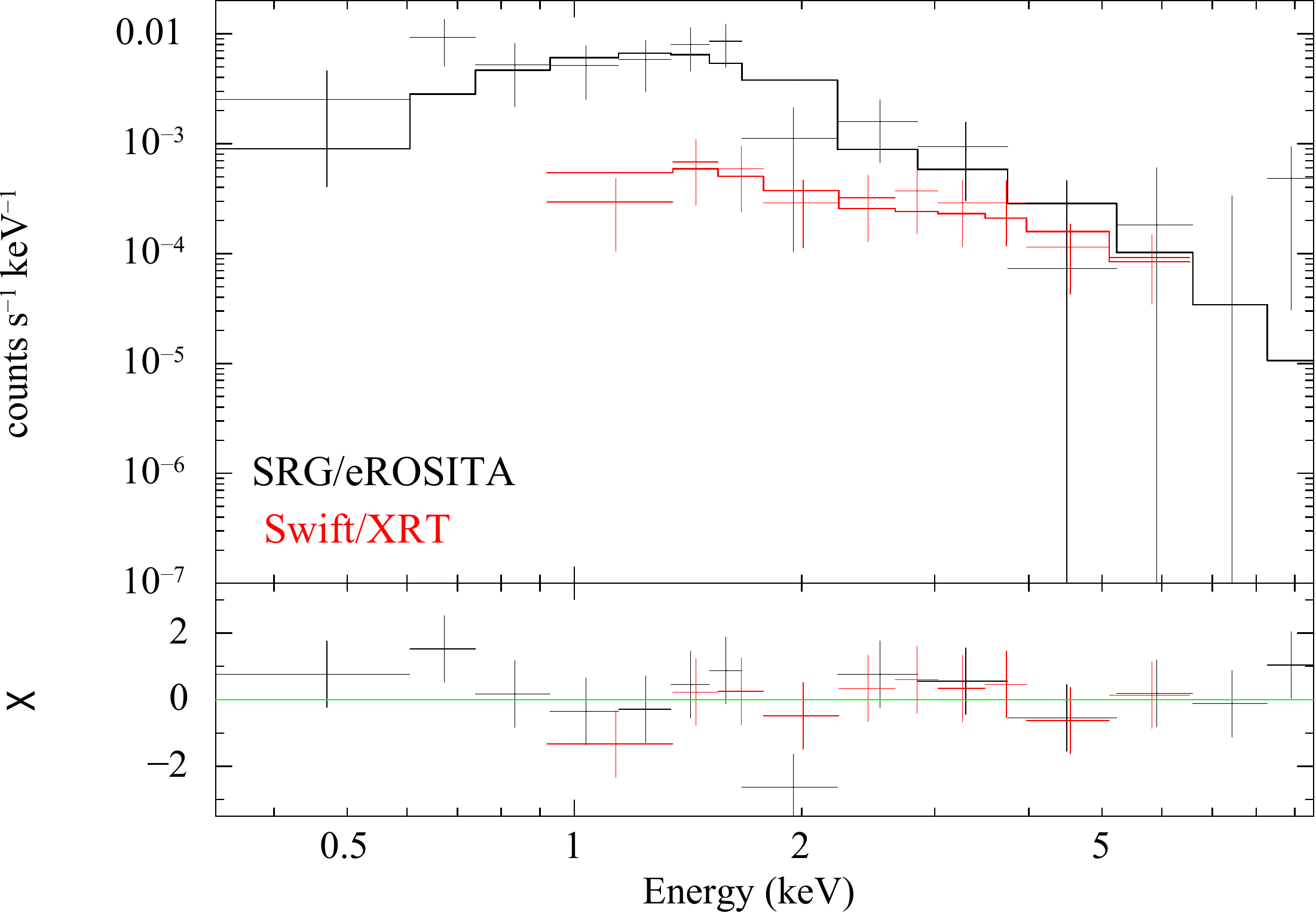}
\caption{The X-ray spectrum of the \src\ counterpart candidate
with the best-fitting PL model (top) and residuals (bottom).
The data obtained by different instruments are marked by different colours
as indicated in the top panel. For illustrative purposes, the \swift\
and \eros\ spectra were regrouped to ensure at least 3 and 10 counts
per energy bin, respectively.}  
    \label{fig:xray-spectrum}
\end{figure}

We fitted the two spectra in the 0.3--10 keV band with the X-Ray Spectral Fitting Package ({\sc xspec}) v.12.11.1 \citep{xspec} 
applying the absorbed power law (PL) model. The {\sc tbabs} model
with the {\sc wilm} abundances \citep*{wilms2000} was used to account
for the interstellar absorption. We transformed  
the reddening $E(B-V)$ found above to the absorbing column density 
$N_{\rm H}=3.3\times10^{21}$~cm$^{-2}$ utilising the relation from 
\citet{foight2016}. We fixed this value during the fitting procedure.
Due to the low number of counts we used the $C$-statistics \citep{cash1979}.
We found that the best-fitting parameters obtained for \eros\ and \swift\
spectra are in agreement within uncertainties indicating that there is 
no statistically significant variability of the source flux 
between \swift\ and \eros\ observations. Therefore, as a next step
we fitted the spectra simultaneously and obtained the photon index $\Gamma = 1.0\pm0.3$,
the unabsorbed flux in the 0.5--10 keV band 
$F_X=1.7^{+0.5}_{-0.3}\times 10^{-13}$ \flux\ (uncertainties correspond 
to 1$\sigma$ confidence intervals) and $C=86$ per 77 degrees of freedom. 
The spectra and the best-fitting model are shown in Fig.~\ref{fig:xray-spectrum}.

\section{Discussion and conclusions}
\label{sec:discussion}

\begin{figure}
\includegraphics[width=\columnwidth]{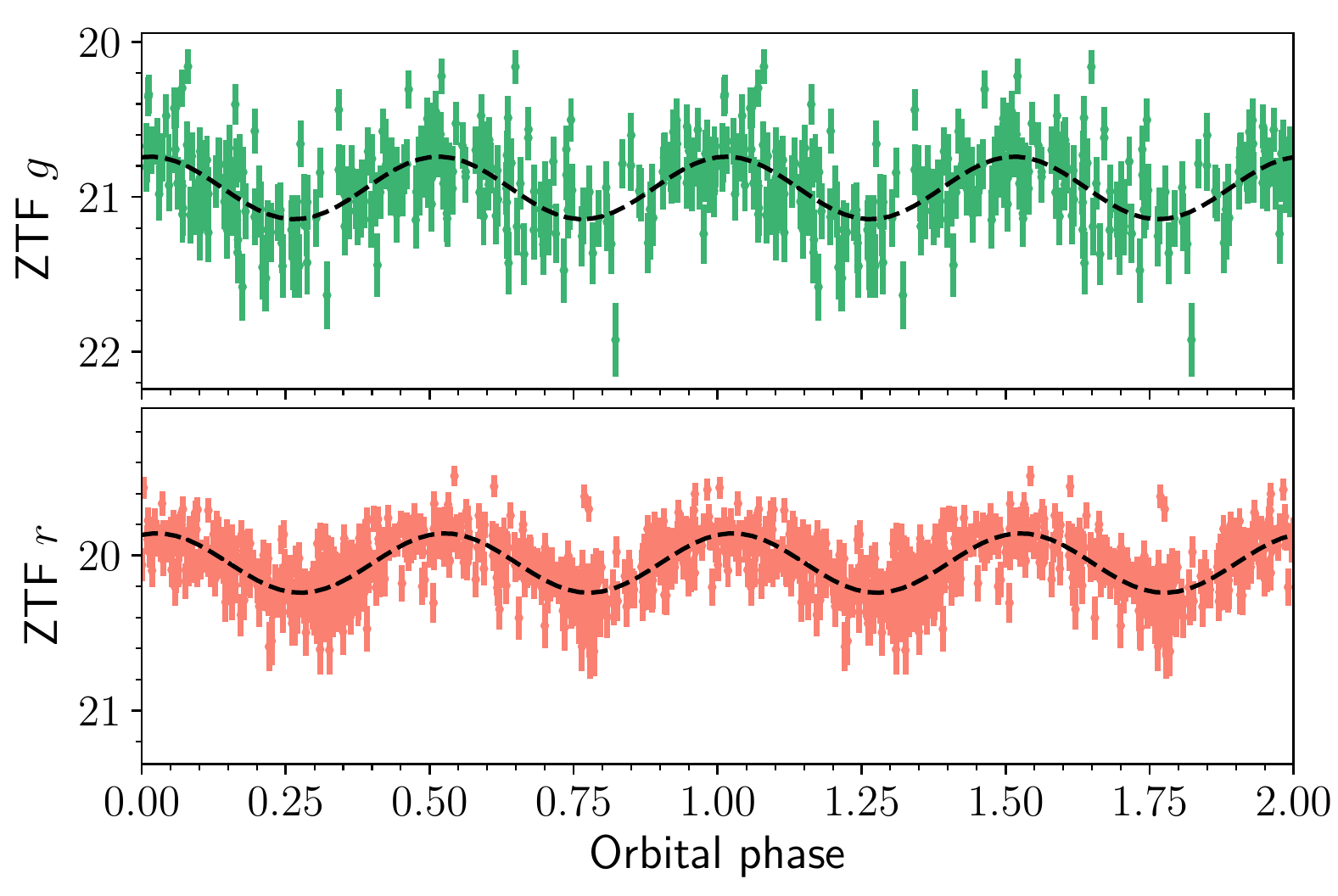}
\caption{ Light curves of the \src\ optical counterpart candidate 
obtained using the ZTF data in the $g$ (\textit{top}) and $r$ bands (\textit{bottom}) folded with the period of $2P_{\rm ph}\approx15$ h. The best-fitting sine curves are shown by the black dashed lines.  }
    \label{fig:lc-15h}
\end{figure}

We examined the nature of the possible X-ray and optical counterpart of 
the unassociated \fermi\ source 4FGL J2054.2$+$6904. We found that 
the source is variable in the optical with a likely 
period 
of about 7.5 h  (Fig.~\ref{fig:ztf-period}). This value lies within a range of orbital 
periods measured for spider systems. The small peak-to-peak amplitude of the light curves
$\approx$0.4~mag (Fig.~\ref{fig:lc})  suggests that the source might belong to the RB class 
whose brightness variations typically do not exceed 1 mag.
In contrast, most BWs have peak-to-peak amplitudes of 2--4 mag \citep[e.g.][]{draghis2019}.
On the other hand, light curves of the majority of RBs show two peaks
per orbital period as a consequence of an ellipsoidal shape of the star due to its tidal distortion.
However, in some RB systems as well as in BW ones, the heating of the companion 
face side by the pulsar wind dominates over the tidal effect
which leads to single-peaked light curves 
\citep{strader2019,swihart2019,swihart2022,kandel2020,linares2018}.
 
Domination of the tidal effect in  \src\ implies   
that the true orbital period can be  twice larger than the derived one, 
$P_b = 2P_{\rm ph}\approx15$ h. In this case, the orbital light curves can have  two peaks and show no colour modulation, 
reflecting an ellipsoidal shape of the star.  
 The observed $g-r$ colour curve appears to support this interpretation,  while the colour  uncertainties are too large to completely  exclude its modulation (Fig.~\ref{fig:lc}, bottom).   
At the same time, including  
two Fourier terms in the Lomb-Scargle algorithm, we indeed found 
two harmonics in the periodogram but the highest peak remained at 7.5 h. 
Typically,  double-peaked light curves of RBs  have two unequal minima due 
to contribution of the heating effect.  
We thus tried to fold the \src\ light curves with $P_b = 2P_{\rm ph}$
and did not find any evidence of this feature (see Fig.~\ref{fig:lc-15h}). 
However, it can be hidden in the noisy data since
the difference between the companion day-side and night-side
temperatures can be rather low, $\lesssim$100--200 K, 
as was derived for, e.g.,  PSRs J1431$-$4715 and J2129$-$0429 
\citep{strader2019,bellm2016}.
We also did not reveal any difference between the shapes of the two peaks due to, e.g.,  hot spots or off-centre heating by the pulsar wind, which would be a strong argument for the 15 h period. 
In addition, the observed variations of RBs brightness due to
the ellipsoidal modulation are usually very low, $\lesssim$0.3 mag
\citep{kaluzny2003,bellm2016,shahbaz2017,sanpaarsaphd,strader2019}, while \src\ shows stronger 
variability of $\approx$0.4 mag which might  indicate  significance of the  irradiation effect. 
It is impossible to make any definite conclusions about that at current data stage. 
To  distinguish the true orbital period of \src, one need either the radio, X-ray or radial velocity data. 
Supplemented with a higher quality photometry, this  would also solve the irradiation dominated vs 
ellipsoidal modulation problem discussed  above.   

The SED (Fig.~\ref{fig:sed}) and the estimated best-fitting effective temperature 
corresponds  to a  
G2--G9-type  
star. Its radius of 
0.5$\pm$0.1~\rsun\ 
is too large
for the very-low mass BW companions, whose  typical radii  are in the range
of $\sim 0.1$~\rsun\ \citep{zharikov2019}. This supports the RB
interpretation of the source.

The X-ray spectrum of the \src\ counterpart candidate can be well
described by the PL model which is typical for spider pulsars
\citep{alnoori2018,strader2019,swihart2022}.
The X-ray and \gr\ luminosities of the source are 
$L_X = 2.0^{+0.6}_{-0.4}\times10^{31}\ (D/{\rm 1\ kpc})^2$ \ergs\
and $L_\gamma = (5.0\pm0.6)\times10^{32}\ (D/{\rm 1\ kpc})^2$ \ergs.
For the accepted distance range of 1.5--5.3 kpc provided by \gaia,  
$L_X = (0.4-7.4)\times10^{32}$ \ergs\ and 
$L_\gamma = (0.1-1.6)\times10^{34}$ \ergs.
The latter value is similar to those observed either for BW or RB pulsars, 
while the former better agrees with the RB interpretation of \src\ 
\citep{strader2019,swihart2022}. The photon index $\Gamma=1.0\pm0.3$ 
is also more typical for RBs than for BWs.

To conclude, according to \gr, X-ray and optical data 
\src\ is a promising RB candidate. 
Its relative optical and X-ray brightness  makes it a good target for further studies.  
Knowledge of the binary period would be helpful in further searches for pulsations with the period of the presumed pulsar in the radio, X- and $\gamma$-rays.  
Optical spectroscopy would allow one to measure its radial velocity curve and to better constrain the spectral type  of the companion star 
and other parameters of the system. Modelling of  multi-band brightness and radial velocity variations with the orbital period  would provide 
fundamental parameters of  \src\, including masses of the binary components, 
the companion temperature distribution over its  surface and the irradiation efficiency 
by the pulsar wind, the distance to the suggested binary system and its inclination.
New dedicated X-ray observations could reveal the orbital modulation of the X-ray emission  
which is often observed for spider pulsars and allow one to constrain the system properties.

\section*{Acknowledgements}

The Pan-STARRS1 Surveys (PS1) and the PS1 public science archive have been made possible 
through contributions by the Institute for Astronomy, the University of Hawaii, 
the Pan-STARRS Project Office, the Max-Planck Society and its participating institutes, 
the Max Planck Institute for Astronomy, Heidelberg and the Max Planck Institute 
for Extraterrestrial Physics, Garching, The Johns Hopkins University, Durham University, 
the University of Edinburgh, the Queen's University Belfast, 
the Harvard-Smithsonian Center for Astrophysics, 
the Las Cumbres Observatory Global Telescope Network Incorporated, 
the National Central University of Taiwan, the Space Telescope Science Institute, 
the National Aeronautics and Space Administration under Grant No. NNX08AR22G 
issued through the Planetary Science Division of the NASA Science Mission Directorate, 
the National Science Foundation Grant No. AST-1238877, the University of Maryland, 
Eotvos Lorand University (ELTE), the Los Alamos National Laboratory, 
and the Gordon and Betty Moore Foundation.

This work is based on observations with eROSITA telescope onboard SRG observatory. The SRG observatory was built by Roskosmos in the interests of the Russian Academy of Sciences represented by its Space Research Institute (IKI) in the framework of the Russian Federal Space Program, with the participation of the Deutsches Zentrum für Luft- und Raumfahrt (DLR). The SRG/eROSITA X-ray telescope was built by a consortium of German Institutes led by MPE, and supported by DLR.  The SRG spacecraft was designed, built, launched and is operated by the Lavochkin Association and its subcontractors. The science data are downlinked via the Deep Space Network Antennae in Bear Lakes, Ussurijsk, and Baykonur, funded by Roskosmos. The eROSITA data used in this work were processed using the eSASS software system developed by the German eROSITA consortium and proprietary data reduction and analysis software developed by the Russian eROSITA Consortium.

DAZ thanks Pirinem School of Theoretical Physics for hospitality.
The work of DAZ and AVK was supported by the Russian Science Foundation  
project 22-12-00048, \url{https://rscf.ru/project/22-12-00048/}.

The authors are grateful to anonymous referee for useful and constructive comments.

\section*{Data Availability}
The \swift/XRT data are available through the archive \url{https://www.swift.ac.uk/swift_portal/}, ZTF data --  \url{https://irsa.ipac.caltech.edu/Missions/ztf.html} and \eros\ data -- on request.




\bibliographystyle{mnras}
\bibliography{ref} 





\bsp	
\label{lastpage}
\end{document}